\documentclass[aps,prl,reprint,groupedaddress]{revtex4-1}
\usepackage{graphicx} 
\usepackage{dcolumn} 
\usepackage{bm}
\usepackage{hyperref}
\usepackage{amsmath}
\usepackage{amssymb}
\usepackage{xcolor}
\usepackage{diagbox}
\usepackage{multirow}
\usepackage{booktabs}
\begin{document}

\title{ Systematic investigation on the superheavy nucleus formation in the reactions of $^{48}$Ca, $^{50}$Ti, $^{51}$V and $^{54}$Cr on actinide nuclei }
\author{Zi-Han Wang$^{1}$ }
\author{Peng-Hui Chen$^{2}$ }
\author{Ya-Ling Zhang$^{3}$ }
\email{Corresponding author: zhangyl@impcas.ac.cn }
\author{Ming-Hui Huang$^{3}$}
\author{Zhao-Qing Feng$^{1,3}$ }
\email{ Corresponding author: fengzhq@scut.edu.cn }
\affiliation{ $^{1}$School of Physics and Optoelectronics, South China University of Technology, Guangzhou 510640, China   \\
$^{2}$School of Physical Science and Technology, Yangzhou University, Yangzhou 225009, China   \\
$^{3}$State Key Laboratory of Heavy Ion Science and Technology, Institute of Modern Physics, Chinese Academy of Sciences, Lanzhou 730000, China
}

\date{\today}

\begin{abstract}
The synthesis of superheavy elements strongly relies on the competition of the quasifission and fusion fission dynamics in the fusion-evaporation reactions. The systematics on the formation of superheavy nuclei in the $^{48}$Ca, $^{50}$Ti, $^{51}$V and $^{54}$Cr induced fusion reactions on actinide nuclei $^{232}$Th, $^{231}$Pa, $^{238}$U, $^{237}$Np, $^{242,244}$Pu, $^{243}$Am, $^{245,248}$Cm, $^{249}$Bk, $^{249}$Cf has been thoroughly investigated with the dinuclear system model by including the cluster transfer and coupling to the dynamical evolution of the quadrupole deformation parameters. The uncertainties of the fusion-evaporation excitation functions with the mass models of FRDM2012, KTUY05, LDM1966, SkyHFB, WS4 are investigated and compared with the available experimental data from Dubna, GSI, Berkeley and RIKEN. The production cross sections, optimal evaporation channels and beam energies in the synthesis of superheavy elements Z = 119 and 120 were predicted and compared for the different mass models in the reactions of $^{50}\mathrm{Ti} + ^{249}\mathrm{Bk}$, $^{51}\mathrm{V} + ^{248}\mathrm{Cm}$, $^{54}\mathrm{Cr} + ^{243}\mathrm{Am}$, $^{50}\mathrm{Ti} + ^{249}\mathrm{Cf}$, $^{51}\mathrm{V} + ^{249}\mathrm{Bk}$, $^{54}\mathrm{Cr} + ^{248}\mathrm{Cm}$, respectively.

\begin{description}
\item[PACS number(s)]
25.70.Hi, 25.70.Lm, 24.60.-k      
\end{description}
\end{abstract}

\maketitle

\section{I. Introduction}
The synthesis of superheavy nuclei (SHN) is one of topic issues in nuclear physics, and associated with the extension of chemical periodic table, the mass limit of atomic nucleus, quantum electrodynamics under the extremely strong Coulomb field \cite{Ke2005}.
Generally, heavy nuclei and superheavy nuclei are relatively unstable and undergo complex interactions, thus evolving towards more stable states. The area where these long-lived and relatively stable SHN are located is called ``island of stability''.
As early as the 1960s, based on the quantum shell effect, Myers and Swiatecki predicted the existence of a ``island of stability" near the proton number Z = 114 and the neutron number N = 184 \cite{My1966}. Based on this prediction, efforts to identify superheavy elements (SHE) in nature were initiated as early as the late 1960s. However, these endeavors were unsuccessful.
With the construction of heavy-ion accelerators and the development of radioactive beams, nuclear physicists have increasingly focused on synthesizing new SHN in the laboratory and have achieved remarkable results in recent years. Currently, the primary experimental methods for producing new isotopes include the reaction types as the nuclear decay, fission and spallation reactions; multiple fragmentation reactions of atomic nuclei; multinucleon transfer (MNT) reactions and heavy-ion fusion-evaporation (FE) reactions. However, The nuclear fragmentation reaction is only suitable for generating nuclides that are lighter than the projectile, and the heaviest stable nuclide currently available is $^{238}$U, so it is powerless to produce nuclides that are heavier than $^{238}$U.
The fission reaction of heavy nuclei is mainly used to produce neutron-rich nuclides in medium mass regions, but not in heavy and superheavy mass regions. The FE reaction is a traditional method for synthesizing new SHN, and in the past half century, experimentalists have successfully synthesized 15 superheavy elements with Z = 103$\sim$118 by FE reactions \cite{Th2013,Th2013(2)}. It is noteworthy that in 2024, at the Lawrence Berkeley National Laboratory, scientists successfully synthesized SHN near the ``island of stability" for the first time via FE reactions using a $^{50}$Ti beam \cite{Ga2024}, providing a reference for exploring elements with Z $>$ 118. Despite this, since the projectile and target nuclei are mostly located near the stability line, the SHE synthesized through the complete fusion reactions are generally neutron-deficient.
Therefore, people began to focus on the study of MNT reactions. Since the 1970s, MNT reactions have been widely studied in experiments \cite{Ad2020}. Particularly, in 2023, the High Energy Accelerator Research Organization (KEK) in Japan discovered a new neutron-rich nuclide, $^{241}$U, through the MNT reaction of $^{238}$U+$^{198}$Pt \cite{Ni2023}. It has further strengthened people's confidence in synthesizing neutron-rich nuclides in heavy and superheavy mass regions through MNT reactions. It is gratifying that over the years, China has also made significant progress in the synthesis of SHE. Recently, at the Heavy Ion Research Facility in Lanzhou (HIRFL), researchers observed an extremely neutron-deficient isotope of $^{210}$Pa through a fusion evaporation reaction of $^{175}$Lu($^{40}$Ca, 5n)$^{210}$Pa \cite{Zhang2024,Zhang2025}. Up to now, the scientists at Institute of Modern Physics (IMP) in Lanzhou has synthesized 41 new isotopes \cite{Gan2024}.

Up to now, the maximum atomic number in the periodic table of elements is oganesson (Z = 118), and finding new elements beyond Og is faced with great difficulties and challenges. Researchers at Dubna made their first attempt to synthesize the element Z = 120 using a reaction system of $^{50}$Fe + $^{244}$Pu \cite{Yu2009}. However, due to the experimental conditions and technical limitations at that time, they failed to observe the corresponding $\alpha$ decay chain. In 2016, GSI proposed using the reaction of $^{54}$Cr + $^{248}$Cm to synthesize Z = 120 and observed three $\alpha$ decay chains \cite{Ho2016}, but subsequent studies indicated that the results might be an accidental phenomenon caused by random events \cite{He2017}. In addition, to create new elements Z = 119 and 120, the reactions of $^{50}$Ti + $^{249}$Bk and $^{249}$Cf were performed at GSI \cite{Kh2020}, and the experiments for $^{51}$V + $^{248}$Cm at RIKEN \cite{Ta2022}. But no confident events for the new elements was obtained in different laboratories.
Therefore, it is essential to accurately predict the optimal reaction systems and energies for the synthesis of SHN from a theoretical perspective, thereby facilitating experimental progress. It is also gratifying that when calculating the production cross-sections of SHN, different theoretical models have all achieved good results, such as the two-step model \cite{Liu2016}, the improved quantum molecular dynamics (ImQMD) model \cite{Wang2002,Zhao2015,Zhao2022}, the multidimensional Langevin equations \cite{Za2008,Za2011,Za2013}, the time-dependent Hartree-Fock (TDHF) method \cite{Si2010,Se2016,Wu2022}, and the GRAZING model \cite{Wi1994,Wi1995,Wen2019}, etc. However, the calculated results have a strong model dependence and our work is carried out through the dinucler system (DNS) model. For the deep inelastic collision, the DNS model \cite{An1995,Ad1997,Feng2007} holds that the incident nucleus is captured by the target nucleus and forming a composite system. During this process, there will be nucleon transfer, the dissipation of energy and angular momentum between the projectile and target, but the two nuclei maintain their respective independence. The DNS model argues that a complete fusion reaction is a special kind of deep inelastic collision. That is, a complete fusion reaction occurs only when all the nucleons in the projectile have transferred to the target. Otherwise, quasi-fission will occur. In the DNS model, the nuclear mass is one of the most fundamental input physical quantities. But so far, only about 3000 atomic nuclei' masses have been accurately measured by experiments. Although the existing mass models can fit the available data relatively accurately, their predictions are significantly different in the unknown mass region. The synthesis cross-sections of SHE are highly sensitive to the input of nuclear masses and the error of the cross-sections given by different mass models may vary by an order of magnitude \cite{Geng2024}.

In this work, the influence of different mass models on the FE excitation functions and the cluster transfer in the synthesis of SHE are to be investigated. The mass tables that we have selected here are respectively the finite-range droplet model (FRDM2012) \cite{Ko2005}, the Koura-Tachibana-Uno-Yamada (KTUY05) \cite{Ko2005}, the little-droplet model (LDM1966) \cite{My1966}, the Hartree-Fock-Bogoliubov method (SkyHFB) \cite{Ba2017} and the Weizs\"{a}cker-Skyrme mass model (WS4) \cite{Wang2014}.
The paper is organized as follows. Section II is a brief introduction of the DNS model that describes the synthesis of SHE. The excitation functions of different systems (the projectile nuclei are respectively $^{48}$Ca, $^{50}$Ti, $^{51}$V, $^{54}$Cr and the target nuclei are respectively $^{232}$Th, $^{231}$Pa, $^{238}$U, $^{237}$Np, $^{242,244}$Pu, $^{243}$Am, $^{245,248}$Cm, $^{249}$Bk, $^{249}$Cf) with different mass tables are systematically analyzed in Sec. III. The conclusions and perspective on the synthesis of new elements with the optimal ways are discussed in Sec. IV.

\section{II. Model Description}
The DNS model describes the reaction mechanism of low energy heavy ion collision as three stages, namely capture, fusion and survival.
In the capture stage, the two colliding nuclei overcome the Coulomb barrier to form a composite system. During the fusion stage, the dissipation of kinetic energy and angular momentum enables the transfer of nucleons between the projectile and target. When all the nucleons are transferred from the lighter nucleus to the heavier nucleus, a composite nucleus (CN) with certain excitation energy and angular momentum can be formed. During the survival stage, the highly excited CN will become the desired nucleus by evaporating light particles such as neutrons, protons, and $\alpha$ particles for deexcitation.
Thus, the production cross-section of the FE reaction can be written as \cite{Feng2007}
\begin{eqnarray}
\sigma _{\rm ER}\left ( E_{\rm c.m.} \right ) &&=\frac{\pi \hbar ^2}{2\mu E_{\rm c.m.}} \sum_{J=0}^{J_{\rm max}}(2J+1)T(E_{\rm c.m.},J)       \nonumber \\
&&\times P_{\rm CN}(E_{\rm c.m.},J)W_{\rm sur}(E_{\rm c.m.},J).
\end{eqnarray}
Here, $E_{c.m.}$ is the incident energy in the centroid coordinate system, $\mu$ is the reduced mass of the collision system, and $J$ represents the orbital angular momentum of the relative motion. We set the maximum angular momentum as $J_{max} = 30 \hbar$. The penetration probability $T(E_{\rm c.m.},J)$ is the probability that the projectile and target nuclei overcome the Coulomb barrier to form a composite system. $P_{CN}(E_{\rm c.m.},J)$ is the fusion probability, that is the formation probability of compound nuclei. The survival probability $W_{sur}(E_{\rm c.m.},J)$ is the probability of the highly excited compound nuclei surviving by evaporating light particles without fission.

\subsection{A. Capture probability}
In the capture stage, when the projectile nucleus and the target nucleus gradually approach each other from infinity, both Coulomb interaction and nuclear interaction occur simultaneously. The calculation formula for the capture cross-section is
\begin{eqnarray}
\sigma_{cap}(E_{c.m.}) &&=  \pi \overline{\lambda}^{2} \sum^{J_{max}}_{J=0} (2J+1)      \nonumber \\
&& \times \int f(B) T(E_{c.m.},J,B) dB,
\end{eqnarray}
and $\lambda = \hbar / \sqrt{2 \mu E_{c.m.}}$ represents the de Broglie wavelength.
Due to the effect of multidimensional quantum penetration (degrees of freedom such as vibration, rotation and deformation), the Coulomb barrier appears a distribution. By introducing a barrier distribution function based on the original Hill-Wheeler formula \cite{Hi1953}, the penetration probability can be written as
\begin{eqnarray}
&& T(E_{\rm c.m.},J)= \nonumber \int   f(B) \\&& \frac{1}{1+\exp\left \{ -\frac{2\pi }{\hbar \omega (J)}\left [ E_{\rm c.m.}-B-\frac{\hbar^2J(J+1)}{2\mu R\rm_B^2(J) }  \right ]  \right \} } dB,
\end{eqnarray}
with $\hbar \omega (J)$ being the width of the parabolic barrier at $R_{B} (J)$. The reduced mass is $\mu = m_{n} A_{P} A_{T} /(A_{P} + A_{T})$ with $m_{n}$, $A_{P}$ and $A_{T}$ being the nucleon mass and the mass numbers of projectile and target nuclei, respectively. $R_{C}$ denotes the Coulomb radius, and $R_{C} = r_{0c}\times(A_{P}^{1/3} + A_{T}^{1/3})$ with $r_{0c} = 1.4\sim1.5$ fm.
The barrier distribution function is Gaussian form \cite{Za2001,Za2001(2)}
\begin{eqnarray}
f(B)= \frac{1}{N} \exp[-((B-B_{m})/\Delta)^{2}].
\end{eqnarray}
The normalization constant satisfies $\int f(B)dB=1$. The quantities $B_{m}$ and $\Delta$ are evaluated by $B_{m}=(B_{C} + B_{S})/2$ and $\Delta = (B_{C} - B_{S})/2$, respectively. The $B_{C}$ is the Coulomb barrier at waist-to-waist orientation and $B_{S}$ is the minimum barrier by varying the quadrupole deformation of the colliding partners. Here we take $B_{S}$ as the Coulomb barrier at tip-to-tip orientation.

\subsection{B. Fusion probability}

Based on the assumption of the DNS model, after a projectile nucleus is captured by a target nucleus, due to the transfer of nucleons and the dissipation of energy and angular momentum, the projectile (or target) at this time is no longer in its original state and is called a projectile-like (or target-like) nucleus. Throughout the entire reaction process, the distribution probability of the DNS fragment is obtained by numerically solving a set of master equation \cite{Ay1976,Ay1976(2),Li2003,Feng2023}.
Fragment ($Z_{1}$,$N_{1}$) has the proton number of $Z_{1}$, the neutron number of $N_{1}$, the internal excitation energy of $E_{1}$, and the quadrupole deformation $\beta_{1}$ and the time evolution equation of its distribution probability can be described as
\begin{widetext}
\begin{eqnarray}
&&  \frac{d P(Z_{1},N_{1},E_{1},\beta_{1},B,t)}{dt}     \nonumber\\
&&  = \sum_{Z^{'}_{1} = Z_{1} \pm 1}W^{p}_{Z_{1},N_{1},\beta_{1};Z^{'}_{1},N_{1},\beta^{'}_{1}}(t) \times [d_{Z_{1},N_{1}}P(Z^{'}_{1},N_{1},E^{'}_{1},\beta^{'}_{1},B,t) - d_{Z^{'}_{1},N_{1}}P(Z_{1},N_{1},E_{1},\beta_{1},B,t)] \nonumber\\
&&  + \sum_{N^{'}_{1} = N_{1} \pm 1}W^{n}_{Z_{1},N_{1},\beta_{1};Z_{1},N^{'}_{1},\beta^{'}_{1}}(t) \times [d_{Z_{1},N_{1}}P(Z_{1},N^{'}_{1},E^{'}_{1},\beta^{'}_{1},B,t) - d_{Z_{1},N^{'}_{1}}P(Z_{1},N_{1},E_{1},\beta_{1},B,t)] \nonumber\\
&&  + \sum_{Z^{'}_{1} = Z_{1} \pm 1, N^{'}_{1} = N_{1} \pm 1}W^{d}_{Z_{1},N_{1},\beta_{1};Z_{1},N^{'}_{1},\beta^{'}_{1}}(t) \times [d_{Z_{1},N_{1}}P(Z^{'}_{1},N^{'}_{1},E^{'}_{1},\beta^{'}_{1},B,t) - d_{Z^{'}_{1},N^{'}_{1}}P(Z_{1},N_{1},E_{1},\beta_{1},B,t)] \nonumber\\
&&  + \sum_{Z^{'}_{1} = Z_{1} \pm 1, N^{'}_{1} = N_{1} \pm 2}W^{t}_{Z_{1},N_{1},\beta_{1};Z_{1},N^{'}_{1},\beta^{'}_{1}}(t) \times [d_{Z_{1},N_{1}}P(Z^{'}_{1},N^{'}_{1},E^{'}_{1},\beta^{'}_{1},B,t) - d_{Z^{'}_{1},N^{'}_{1}}P(Z_{1},N_{1},E_{1},\beta_{1},B,t)] \nonumber\\
&&  + \sum_{Z^{'}_{1} = Z_{1} \pm 2, N^{'}_{1} = N_{1} \pm 1}W^{^{3}He}_{Z_{1},N_{1},\beta_{1};Z_{1},N^{'}_{1},\beta^{'}_{1}}(t)  \times [d_{Z_{1},N_{1}}P(Z^{'}_{1},N^{'}_{1},E^{'}_{1},\beta^{'}_{1},B,t) - d_{Z^{'}_{1},N^{'}_{1}}P(Z_{1},N_{1},E_{1},\beta_{1},B,t)] \nonumber\\
&&  + \sum_{Z^{'}_{1} = Z_{1} \pm 2, N^{'}_{1} = N_{1} \pm 2}W^{\alpha}_{Z_{1},N_{1},\beta_{1};Z_{1},N^{'}_{1},\beta^{'}_{1}}(t) \times [d_{Z_{1},N_{1}}P(Z^{'}_{1},N^{'}_{1},E^{'}_{1},\beta^{'}_{1},B,t) - d_{Z^{'}_{1},N^{'}_{1}}P(Z_{1},N_{1},E_{1},\beta_{1},B,t)].
\end{eqnarray}
\end{widetext}
In this equation, $W_{Z_{1},N_{1},\beta_{1};Z_{1}^{\prime},N_{1}^{\prime},\beta_{1}^{\prime}}$ is the mean transition probability from the channel $(Z_{1},N_{1},E_{1},\beta_{1})$ to $(Z_{1}^{\prime},N_{1}^{\prime},E_{1}^{\prime},\beta_{1}^{\prime})$.
The quantity $d_{Z_{1},N_{1}}$ indicates the microscopic dimension corresponding to the macroscopic state $(Z_{1},N_{1},E_{1},\beta_{1})$.
In this process, the transfer of nucleons or clusters is satisfied with the relationships, $Z_{1}^{\prime} = Z_{1} \pm Z_{\nu}$ and $N_{1}^{\prime} = N_{1} \pm N_{\nu}$, each represents the transfer of a neutron, proton, deuteron, tritium, $^{3}$He, $\alpha$.
The initial probabilities of projectile and target nuclei are set to be $P(Z_{proj},N_{proj},E_{1}=0, t=0) = P(Z_{targ},N_{targ},E_{1}=0, t=0)=0.5$. The nucleon transfer process satisfies the unitary condition $\sum_{Z_{1},N_{1}} P(Z_{1}, N_{1}, E_{1}, t)=1$.
The excitation energy $E_{1}$ is the local excitation energy $\varepsilon_{1}^{*}$ of the fragment $(Z_{1},N_{1})$.
The reaction time $t$ is calculated by the deflection function method \cite{Wo1978,Ri1979,Li1981,Li1983}.

The movement of nucleons and clusters is controlled by the single-particle Hamiltonian, $H(t)= H_{0}(t) + V(t)$ \cite{Ay1976,Ay1976(2),Feng2006}.
The total single particle energy is
\begin{eqnarray}
H_0(t) &&= \sum _K\sum_{\nu_K} \varepsilon_{\nu_K}(t)\alpha^+_{\nu_K}(t)\alpha_{\nu_K}(t).
\end{eqnarray}
The interaction potential is
\begin{eqnarray}
V(t) &&= \sum _{K,K^{'}} \sum_{\alpha_K,\beta_{K'}} u_{\alpha_K,\beta_{K'}}(t)\alpha^+_{\alpha_K}(t)\alpha_{\beta_{K'}}(t) \nonumber  \\
&&= \sum_{K,K'}V_{K,K'}(t).
\end{eqnarray}
The quantity $\varepsilon_{\nu K}$ represents the single-particle energies, and $u_{\alpha_K,\beta_{K'}}$ is the interaction matrix elements. The single-particle states are defined with respect to the centers of the interacting nuclei and are assumed to be orthogonalized in the overlap region. Thus, the annihilation and creation operators depend on the reaction time \cite{Feng2006,Li2003,Li2006}.
Besides, the parameterized matrix element is denoted by
\begin{eqnarray}
&& u_{\alpha_K,\beta_K'} =  U_{K,K'}(t)  \\ && \times \left\{ \exp \left[- \frac{1}{2}( \frac{\varepsilon_{\alpha_K}(t) - \varepsilon_{\beta_K}(t)}{\Delta_{K,K'}(t)})^2 \right] - \delta_{\alpha_K,\beta_{K'}} \right\}, \nonumber
\end{eqnarray}
and the interaction intensity factor is expressed as
\begin{eqnarray}
U_{K,K'} =  {\frac{g_{1}^{1/3} \times g_{2}^{1/3}} {g_{1}^{1/3} + g_{2}^{1/3}}}\cdot{\frac{1} {g_{K}^{1/3} \times g_{K'}^{1/3}}}\cdot 2 \gamma_{K,K'},
\end{eqnarray}
with $K, K' = 1, 2$ for the DNS fragments with the parameters $\gamma_{K,K'} = 3$ and $\Delta_{K,K'} = 2$.

The mean transition probability is related with the local excitation energy and the transfer of nucleons and clusters, and it can be microscopically derived from the interaction potential in valence space as
\begin{eqnarray}
&& W^{\nu}_{Z_{1},N_{1};Z_{1}^{\prime},N^{\prime}_{1}} = G_{\nu} \frac{\tau_{mem}(Z_{1},N_{1},E_{1};Z_{1}^{\prime},N_{1}^{\prime},E_{1}^{\prime})}{d_{Z_{1},N_{1}} d_{Z_{1}^{\prime},N_{1}^{\prime}}\hbar^{2}}    \nonumber \\
&& \times \sum_{ii^{\prime}}|\langle  Z_{1}^{\prime},N_{1}^{\prime},E_{1}^{\prime},i^{\prime}|V|Z_{1},N_{1},E_{1},i \rangle|^{2}.
\end{eqnarray}
$G_{\nu}$ represents the spin-isospin statistical factors, and we use the winger density approach to identify particle types \cite{Ma1997,Feng2020}, i.e. $G_{\nu} = 1, 1, 3/8, 1/12, 1/12, 1/96$ for neutron, proton, deuteron, tritium, $^{3}$He, $\alpha$, respectively.
It is noticed that the cluster transition probability is related to the cluster formation probability with the cluster structure, cluster potential, cluster binding energy, Mott effect etc. More improvements are still needed in the future work.
The memory time is connected with the internal excitation energy,
\begin{eqnarray}
&& \tau_{mem}(Z_1,N_1,E_1; Z'_1,N_1, E'_1) =   \nonumber \\
&& \left[\frac{2\pi \hbar^2} {\sum _{KK'} <V_{KK} V^*_{KK'}>}\right]^{1/2},
\end{eqnarray}
\begin{eqnarray}
&& <V_{KK} V^*_{KK'}> = \frac{1}{4} U^2_{KK'}g_K g_K' \Delta_{KK'} \Delta \varepsilon_K \Delta \varepsilon_K^{\prime}
 \nonumber \\
&&  \times \left[ \Delta^2_{KK'}+ \frac{1}{6} ((\Delta \varepsilon_K)^2  + (\Delta \varepsilon_K^{\prime})^2) \right]^{-1/2}.
\end{eqnarray}

The evolution of DNS with the distance R between two nuclei may lead to quasi-fission. During the fusion process, highly excited fragments may lead to fission. In the calculations, we overlooked the quasi-fission of DNS and the fission of heavy fragments.
During the relaxation process of relative motion, the kinetic energy of relative motion of the projectile and target nuclei dissipates into the dinuclear system and becomes the intrinsic excitation energy of the system. The excited dinuclear system opens a valence space in which the valence nucleons have a symmetrical distribution around the Fermi surface. Only the particles at the states within the valence space are actively excited and undergo transfer. The averages on these quantities in the valence space are as follows:
\begin{eqnarray}
\Delta \varepsilon_K = \sqrt{\frac{4\varepsilon^*_K}{g_K}},\quad
\varepsilon^*_K =\varepsilon^*\frac{A_K}{A}, \quad
g_K = A_K /12,
\end{eqnarray}
here, the symbol $\varepsilon^{*}$ is the local excitation energy of the DNS fragment.
The number of valence states in valence space is $N_K$ = $g_{\rm K}\Delta\varepsilon_{\rm K}$, $g_{\rm K}$ is the single particle level density around the Fermi surface.
The number of valence nucleon is $m_{\rm K}$ = $N_{\rm K/2}$.
The microscopic dimension for the fragment ($Z_{K},N_{K}$) is evaluated by
\begin{eqnarray}
 d(m_1, m_2) = {N_1 \choose m_1} {N_2 \choose m_2}.
\end{eqnarray}

The local excitation energy of DNS fragment is related to the dissipated energy and potential energy surface (PES) of the relative motion, which provides the excitation energy of the average transition probability, as given by the following formula
\begin{eqnarray}
\varepsilon^{\ast}(t)=E_{diss}(t)-\left(U(\{\alpha\})-U(\{\alpha_{EN}\})\right).
\end{eqnarray}
where ${\alpha_{EN}}={Z_{P},N_{P},Z_{T},N_{T},J,R,\beta_{P},\beta_{T},\theta_{P},\theta_{T}}$ for the projectile-target system, and J represents the initial angular momentum.
The excitation energy of the DNS fragment$(Z_{1},N_{1})$ is $E_{1}=\varepsilon^{\ast}(t=\tau_{int})A_{1}/A$.
$\tau_{int}$ denotes the interaction time, which is associated with the reaction system and the relative angular momentum, and can be gained by the deflection function.
The dissipated energy of relative motion varies with time \cite{Li1981,Wo1978}:
\begin{equation}
E_{diss}(t)=E_{c.m.}-B-\frac{\langle  J(t)\rangle(\langle J(t)\rangle+1)\hbar^{2}}{2\zeta_{rel}}-\langle  E_{rad}(J,t)\rangle.
\end{equation}
And the radial energy is
\begin{equation}
\langle  E_{rad}(J,t)\rangle=E_{rad}(J,0)\exp(-t/\tau_{r}),
\end{equation}
with the relaxation time of radial motion being $\tau_{r} = 5 \times 10^{-22}$ s, the initial radial energy being $E_{rad}(J,0)=E_{c.m.}-B-J_{i}(J_{i}+1)\hbar^{2}/(2\zeta_{rel})$.
The dissipation of the relative angular momentum is described by
\begin{equation}
\langle  J(t)\rangle=J_{st}+(J_{i}-J_{st})\exp(-t/\tau_{J})
\end{equation}
The angular momentum at the sticking limit is $J_{st}=J_{i}\zeta_{rel}/\zeta_{tot}$ and the relaxation time is $\tau_{J}=15\times10^{-22}$ s.
$\zeta_{rel}$ and $\zeta_{tot}$ are the relative and total moments of inertia of the DNS, respectively.
The initial angular momentum is set to be $J_{i}=J$ in Eq. (1).
The PES of the DNS is evaluated as
\begin{eqnarray}
U(\{\alpha\}) && = B(Z_{1},N_{1}) + B(Z_{2},N_{2})   \nonumber \\
&& - B(Z,N) + V(\{\alpha\}),
\end{eqnarray}
with the relationship of $Z_{1} + Z_{2} = Z$ and $N_{1} + N_{2} = N$.
The symbol $\{\alpha\}$ denotes the quantities ${Z_{1},N_{1},Z_{2},N_{2},J,R,\beta_{1},\beta_{2},\theta_{1},\theta_{2}}$.
In the calculation, the distance between the centers of the two fragments $R$ is chosen to be the value at the touching configuration, in which the DNS is assumed to be formed, and $R = r_{0}*(A_{P}^{1/3} + A_{T}^{1/3})$ with $r_{0} = 1.2\sim1.3$ fm.
The $\beta_{i}$ represent the quadrupole deformations of the two fragments at ground state, and $\theta_{i} (i=1,2)$ denote the angles between the collision orientations and the symmetry axes of deformed nuclei.
The $B(Z_{i},N_{i}) (i=1,2)$ and $B(Z,N)$ are respectively the negative binding energies of the fragment $(Z_{i},N_{i})$ and the compound nucleus $(Z,N)$, which are read from different mass tables.
The interaction potential energy between fragment $(Z_{1},N_{1})$ and $(Z_{2},N_{2})$ is derived from
\begin{eqnarray}
V(\{\alpha\}) = V_{C}(\{\alpha\}) + V_{N}(\{\alpha\}) + V_{def}(t),
\end{eqnarray}
and
\begin{eqnarray}
V_{def}(t) = \frac{1}{2} C_{1} (\beta_{1}^{0} - \beta_{1}^{'}(t))^{2} + \frac{1}{2} C_{2} (\beta_{2}^{0} - \beta_{2}^{'}(t))^{2} .
\end{eqnarray}
Here, $V_{C}$ is the Coulomb potential using the Wong formula \cite{Wong1973}, $V_{N}$ is the nucleus-nucleus potential using the double folding potential \cite{Go2004}, and $V_{def}(t)$ denotes the deformation energy of the DNS at reaction time t.
$\beta_{1,2}^{0}$ are the ground state quadrupole deformations \cite{Mo1995} and $\beta_{1,2}^{'}(t)$ are the dynamic quadrupole deformations of projectile-like  fragment and target-like fragment.
The quantity $C_{1,2}$ denote the stiffness parameters of the nuclear surface, which are calculated by the liquid drop model \cite{My1966}.
Detailed calculations of $V_{def}(t)$ can be obtained from Ref. \cite{Feng2006,Chen2023} and the references therein.

The DNS model typically assumes that at the lowest point of the interaction potential, a dinuclear system can exist for a long time and undergo nucleon transfer, that is to say, nucleon transfer occurs at the bottom of the pocket ($R = R_{min}$).
The position of the B.G. point (the Businaro-Gallone point) represents the position of the highest point of the PES along the direction of mass asymmetry.
And the inner fusion barrier is defined as the difference in driving potential between the B.G. point and the incident point.
These fragments that have overcome the internal fusion barrier, are considered to have undergone fusion.
Therefore, the probability of fusion is given by the following formula \cite{Ad1997,Feng2009},
\begin{eqnarray}
&& P_{CN}(E_{c.m.},B,J)= \nonumber \\ && \sum _{Z_{1}=1}^{Z_{BG}} \sum _{N_{1}=1}^{N_{BG}} P(Z_{1},N_{1},E_{1}(J),\tau_{int}(J)).
\end{eqnarray}
\begin{eqnarray}
P_{CN}(E_{c.m.},J)=\int f(B)P_{CN}(E_{c.m.},B,J)dB
\end{eqnarray}

\subsection{C. Survival probability}
Through the transfer of particles, the projectile and the target form a compound nucleus with a certain excitation energy.
However, the excited state of the composite nucleus is highly unstable and can be deexcited by emitting $\gamma$ rays, evaporating n, p and $\alpha$ particles, or undergoing fission.
This process can be described by the Weisskopf statistical evaporation model \cite{We1937}.
If x represents the number of neutrons emitted, y represents protons, and z represents $\alpha$ particles, then the survival probability can be expressed as
\begin{eqnarray}
&&W_{sur}(E_{CN}^*,x,y,z,J)=P(E_{CN}^*,x,y,z,J) \times \nonumber\\&&  \prod_{i=1}^{x}\frac{\Gamma _n(E_i^*,J)}{\Gamma _{\rm tot}(E_i^*,J)} \prod_{j=1}^{y}\frac{\Gamma _p(E_j^*,J)}{\Gamma_{\rm tot}(E_i^*,J)} \prod_{k=1}^{z}\frac{\Gamma _\alpha (E_k^*,J)}{\Gamma _{\rm tot}(E_k^*,J)},
\end{eqnarray}
where the $E_{CN}^*$ and $J$ are the excitation energy and the spin of the excited nucleus respectively.
The total width $\Gamma_{tot}$ is the sum of partial widths of particle evaporation, $\gamma$-rays and fission.
The excitation energy $E_{s}^*$ before evaporating the $s$-th particles is evaluated by
\begin{eqnarray}
E_{s+1}^*=E_s^* - B _i ^n - B _j ^p - B_k ^\alpha - 2T_s
\end{eqnarray}
with the initial condition $E_{i}^*$ = $E_{CN}^*$ and $s$ = $i$ + $j$ + $k$.
The $B_{i}^n$, $B_{j}^p$, $B_{k}^\alpha$ are the separation energy of the $i$-th neutron, $j$-th proton, $k$-th alpha, respectively.
The nuclear temperature $T_i$ is defined by $E_{i}^* = a T_{i}^2-T_{i}$ with the level density parameter $a$ = $A/8$ \cite{Feng2009,Feng2009(2)}.
For one particle channel, the realization probability of evaporation channels $P(E_{CN}^*,J)$ is written as
\begin{eqnarray}
P(E_{CN}^*,J)= \exp\left ( -\frac{(E_{CN}^*-B_s-2T)^2}{2\sigma ^2}  \right )
\end{eqnarray}
where $\sigma$ is taken as 2.2 $\sim$ 2.5 MeV, which is the half-height width of the excitation function of the residual nuclei.
For the multiple neutrons channels $(x>1)$, the $P(E_{CN}^*,J)$ could be derived by the Jackson formula \cite{Ja1956}.
The $\Gamma _n(E_i^*,J)$, $\Gamma _p(E_i^*,J)$ and $\Gamma _{\alpha}(E_i^*,J)$ are respectively the decay widths of particles n, p and $\alpha$.
In the calculation, we only considered the evaporation of neutrons.
The fission width $\Gamma_f(E^*,J)$ is calculated by the Bohr-Wheeler formula \cite{Bo1939}.

The fission barrier has a microscopic part and a macroscopic part in our work \cite{Ad2000}.
Considering the correlation with angular momentum, it is written as
\begin{eqnarray}
&& B_f(E^*,J)=B_f^{LD} \nonumber \\ && +B_f^M(E^*=0,J) \times exp(-E^*/E_D)
\end{eqnarray}
where the macroscopic part is derived from the liquid-drop model, as follows
\begin{eqnarray}
B^{LD}_f = \left \{\begin{array} {rl} 0.38(0.75 - x )E_{s0}&, (1/3 < x < 2/3) \\ \\
0.83(1-x)^3 E_{s0}&, (2/3 < x < 1) \end{array} \right.
\end{eqnarray}
with
\begin{eqnarray}
x=\frac{E_{c0}}{2E_{s0}}.
\end{eqnarray}
Here, $E_{c0}$ and $E_{s0}$ are the surface energy and Coulomb energy of the spherical nuclear respectively, which could be taken from the Myers-Swiatecki formula \cite{My1974}:
\begin{eqnarray}
E_{s0}=17.944[1-1.7826(\frac{N-Z}{A})^2]A^{2/3} \ MeV
\end{eqnarray}
and
\begin{eqnarray}
E_{c0}=0.7053\frac{Z^2}{A^{1/3}} \ MeV.
\end{eqnarray}
Shell-damping energy is taken as
\begin{eqnarray}
E_{D}=0.4\frac{A^4/3}{a} \ MeV.
\end{eqnarray}

The level density is calculated from the Fermi-gas model as
\begin{eqnarray}
\rho(E^{\ast},J) && = K_{rot} K_{vib} \times \frac{2J+1}{24\sqrt{2}\sigma^3a^{1/4}(E^{\ast} - \delta)^{5/4}}   \nonumber \\
&& \times\exp\left[ 2\sqrt{a(E^{\ast}-\delta)} - \frac{(J+1/2)^2}{2\sigma^2}\right],
\end{eqnarray}
with $\sigma^2 = 6\bar{m}^2\sqrt{a(E^*-\delta)}/\pi^2$ and $\bar{m}\approx0.24A^{2/3}$.
Besides, $K_{rot}$ and $K_{vib}$ are respectively the rotational enhancement coefficient and the vibration enhancement coefficient, and take $K_{rot}$ = $K_{vib}$ = 1.
The pairing correction energy $\delta$ is set to be $12/\sqrt{A}, 0$ and $-12/\sqrt{A}$ MeV for even-even, even-odd and odd-odd nuclei, respectively.
The level density parameter is related to the shell correction energy $E_{sh}(Z,N)$ \cite{Mo1995} and the excitation energy $E^{\ast}$ of the nucleus as
\begin{equation}
a(E^{*},Z,N)=\tilde{a}(A)[1+E_{sh}(Z,N)f(E^{\ast}-\Delta)/(E^{\ast}-\Delta)].
\end{equation}
The asymptotic Fermi-gas value of the level density parameter at high excitation energy is $\tilde{a}(A) = \alpha A + \beta  A^{2/3}b_{s}$ MeV$^{-1}$, and the shell damping factor is given by $f(E^{\ast}) = 1-\exp(-\gamma E^{\ast})$ with $\gamma=\tilde{a}/(\epsilon  A^{4/3})$.
In our calculation, we take $a = \tilde{a}(A)$, and the parameters $\alpha$, $\beta$, $b_{s}$ and $\epsilon$ are taken to be 0.114, 0.098, 1 and 0.4, respectively.

\section{III. Results and discussion}
The DNS concept has been extensively used for the deep inelastic nuclear collisions, quasifission dynamics, $\alpha$ decay and spontaneous fission, fusion-evaporation reactions, multinucleon transfer reactions (MNT) etc \cite{Ad1996,Ad1997, Ad1998,Feng2009(3)}. It has been manifested that the powerful prediction and reasonable explanation of a number of experimental data, i.e., the fusion-evaporation excitation functions, optimal beam energy and evaporation channel for synthesizing the new SHN, angular distribution and kinetic energy spectra of projectile-like or target-like fragments in the MNT reactions \cite{Bao2016, Chen2016, Chen2020, Li2022, Chen2023(2), Le2025, Zhang2025(2)}. However, it is well known that the calculated results of the DNS model depend on the input quantities, such as the nuclear mass, the fission barrier, potential energy surface etc. Variations in these inputs can lead to large differences in the calculated production cross-sections, often by several orders of magnitude.
In this work, using five different mass tables (FRDM2012, KTUY05, LDM1966, SkyHFB, WS4), we systematically calculated the FE reactions induced by $^{48}$Ca, $^{50}$Ti, $^{51}$V and $^{54}$Cr on different actinide nuclides in the framework of our updated DNS model. The Q-values of each reaction system under different mass tables are presented in Table\ref{Tab1}. Specially, the reaction system $^{48}\mathrm{Ca} + ^{248}\mathrm{Cm} \rightarrow ^{296}116^{*}$ was selected as a representative case to investigate the impact of different mass tables on each stage of the DNS model.

\begin{table*}
\centering
\caption{\label{Tab1} Q-values (MeV) calculated using different mass tables (M-Table) for each reaction system, and the P represents the projectile nucleus and T for the target nucleus, $Q = \Delta_{1}(Z_{1},N_{1}) + \Delta_{2}(Z_{2},N_{2}) - \Delta_{CN}(Z_{1}+Z_{2},N_{1}+N_{2})$ in MeV. }
\small
\begin{ruledtabular}
\begin{tabular}{cccccccccccccc}
\hline
\diagbox{P}{T} & M-Table & $^{232}$Th & $^{231}$Pa & $^{238}$U & $^{237}$Np & $^{242}$Pu & $^{244}$Pu & $^{243}$Am & $^{245}$Cm & $^{248}$Cm & $^{249}$Bk & $^{249}$Cf  \\
\hline
\multirow{5}{*}{$^{48}$Ca}
& FRDM2012 & -154.72 & -158.68 & -157.76 & -161.83 & -161.84 & -160.44 & -164.98 & -167.81 & -166.57 & -170.07 & -174.18 \\
& KTUY05 & -160.87 & -164.89 & -163.13 & -167.67 & -166.85 & -164.74 & -169.80 & -171.93 & -169.55 & -172.49 & -176.14 \\
& LDM1966 & 154.08 & -158.00 & -156.83 & -161.01 & -160.67 & -159.34 & -164.06 & -167.20 & -165.65 & -169.20 & -173.33 \\
& SkyHFB & -155.40 & -159.31 & -158.70 & -163.10 & -163.67 & -162.40 & -167.20 & -168.63 & -167.46 & -170.01 & -173.06 \\
& WS4 & -157.00 & -160.878 & -160.40 & -164.54 & -165.12 & -163.70 & -168.24 & -170.93 & -169.49 & -172.87 & -176.67 \\
\hline
\multirow{5}{*}{$^{50}$Ti}
& FRDM2012 & -171.87 & -177.13 & -174.61 & -180.53 &         & -178.85 & -184.74 &         & -186.29 & -190.66 & -195.97 \\
& KTUY05 & -177.12 & -182.68 & -179.22 & -185.24 &         & -181.68 & -187.86 &         & -187.10 & -190.71 & -195.26 \\
& LDM1966 & -170.51 & -175.95 & -173.25 & -179.35 &         & -177.28 & -183.72 &         & -185.01 & -189.36 & -194.72 \\
& SkyHFB & -172.85 & -178.12 & -176.16 & -182.18 &         & -180.09 & -185.75 &         & -185.79 & -189.11 & -193.09 \\
& WS4 & -173.13 & -178.78 & -176.22 & -182.07 &         & -180.34 & -186.51 &         & -187.42 & -191.66 & -196.66 \\
\hline
\multirow{5}{*}{$^{51}$V}
& FRDM2012 & -177.94 & -182.01 & -181.31 & -186.47 &         & -185.83 & -191.36 &         & -193.86 & -197.29 &         \\
& KTUY05 & -183.75 & -187.99 & -186.10 & -190.71 &         & -188.68 & -193.52 &         & -194.46 & -196.51 &         \\
& LDM1966 & -175.95 & -181.09 & -179.25 & -185.31 &         & -183.87 & -190.03 &         & -191.91 & -196.05 &         \\
& SkyHFB & -179.57 & -184.05 & -183.34 & -188.17 &         & -186.96 & -191.86 &         & -192.95 & -195.33 &         \\
& WS4 & -179.71 & -184.24 & -183.00 & -187.78 &         & -187.41 & -192.45 &         & -195.07 & -197.82 &         \\
\hline
\multirow{5}{*}{$^{54}$Cr}
& FRDM2012 & -188.35 & -194.44 & -192.44 & -199.29 &         & -198.35 & -204.65 &         & -206.85 &         &         \\
& KTUY05 & -194.29 & -200.81 & -197.00 & -203.58 &         & -200.24 & -206.59 &         & -206.34 &         &         \\
& LDM1966 & -187.66 & -194.05 & -191.84 & -198.86 &         & -197.56 & -204.27 &         & -206.52 &         &         \\
& SkyHFB & -192.03 & -198.06 & -195.71 & -202.21 &         & -200.08 & -206.30 &         & -206.18 &         &         \\
& WS4 & -191.27 & -197.67 & -195.51 & -201.95 &         & -200.69 & -207.10 &         & -208.48 &         &         \\
\hline
\end{tabular}
\end{ruledtabular}
\end{table*}

In the DNS model, it is generally posited that a projectile nucleus with a specific incident energy, overcomes the Coulomb barrier to form a dinuclear system. Subsequently, nucleon transfer occurs at the pocket of the nuclear-nuclear interaction potential. Therefore, the PES only depends on the degree of mass asymmetry, namely the driving potential. Fig. 1 shows the driving potential energy of the reaction of  $^{48}\mathrm{Ca} + ^{248}\mathrm{Cm}$ with the five mass tables. In the figure, the black arrow indicates the position of the incident point ($\eta=-0.6757$), the five-pointed stars represent the Businaro-Gallone (B.G.) points of different mass tables, and among them, the red one represents FRDM2012 ($\eta=-0.8176$), the blue one is KTUY05 ($\eta=-0.8446$), the magenta is LDM1966 ($\eta=-0.8176$), the dark yellow is SkyHFB ($\eta=-0.8581$) and dark cyan represents WS4 ($\eta=-0.8851$). And the corresponding internal fusion barriers are $B_{fus}(FRDM2012)=14.9946$ MeV, $B_{fus}(KTUY05)=17.2527$ MeV, $B_{fus}(LDM1966)=15.2321$ MeV, $B_{fus}(SkyHFB)=13.8170$ MeV and $B_{fus}(WS4)=15.4945$ MeV, respectively. It is obvious that the mass model SkyHFB exhibits the smallest internal fusion barrier, followed by the FRMD2012 model, and the absolute value of $\eta$ at B.G. point of FRMD2012 is the smallest. The smaller the internal fusion barrier is, the more conducive it is to the occurrence of fusion events. The closer the B.G. point is to the incident position, the more larger fusion probability.

\begin{figure}
\centering
\includegraphics[width=8 cm]{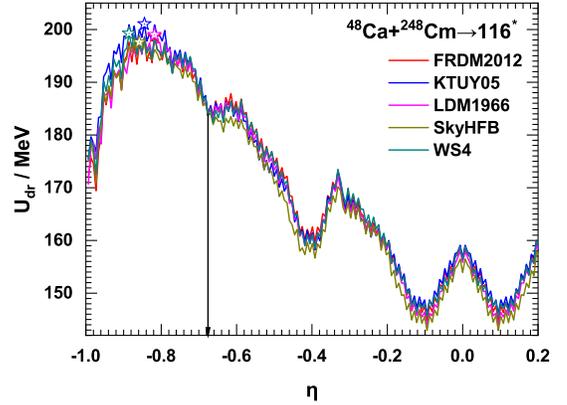}
\caption{Driving potential of the reaction of $^{48}\mathrm{Ca} + ^{248}\mathrm{Cm}$ with different mass tables. The black arrow indicates the position of the incident point. The five-pointed stars represent the B.G. points of different mass tables, and the red line for FRDM2012, blue line for KTUY05, magenta line for LDM1966, yellow line for SkyHFB and dark cyan line for WS4, respectively.}
\end{figure}

\begin{figure*}
\centering
\includegraphics[width=16 cm]{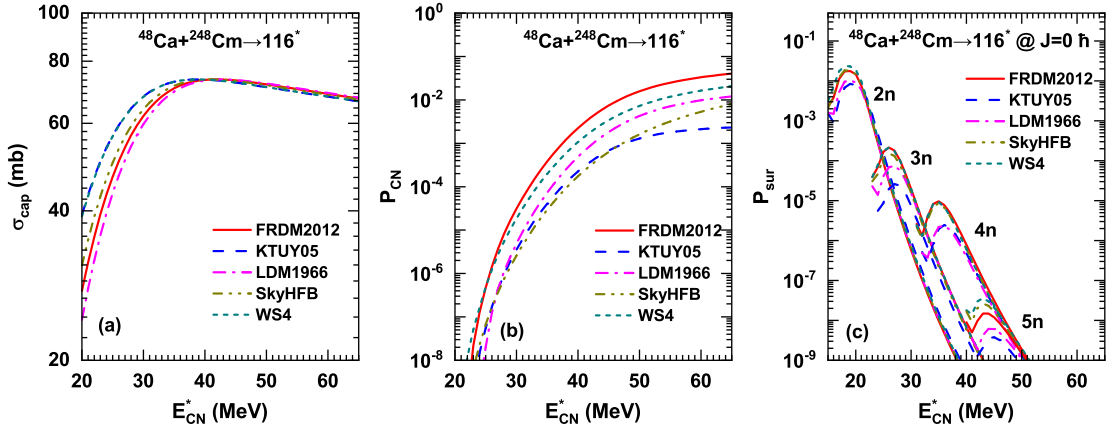}
\caption{ (a) The capture cross section, (b) fusion probability and (c) survival probability as a function of the excitation energy of compound nucleus in the reaction of $^{48}\mathrm{Ca} + ^{248}\mathrm{Cm}$ with different mass models.}
\end{figure*}

\begin{figure*}
\centering
\includegraphics[width=16 cm]{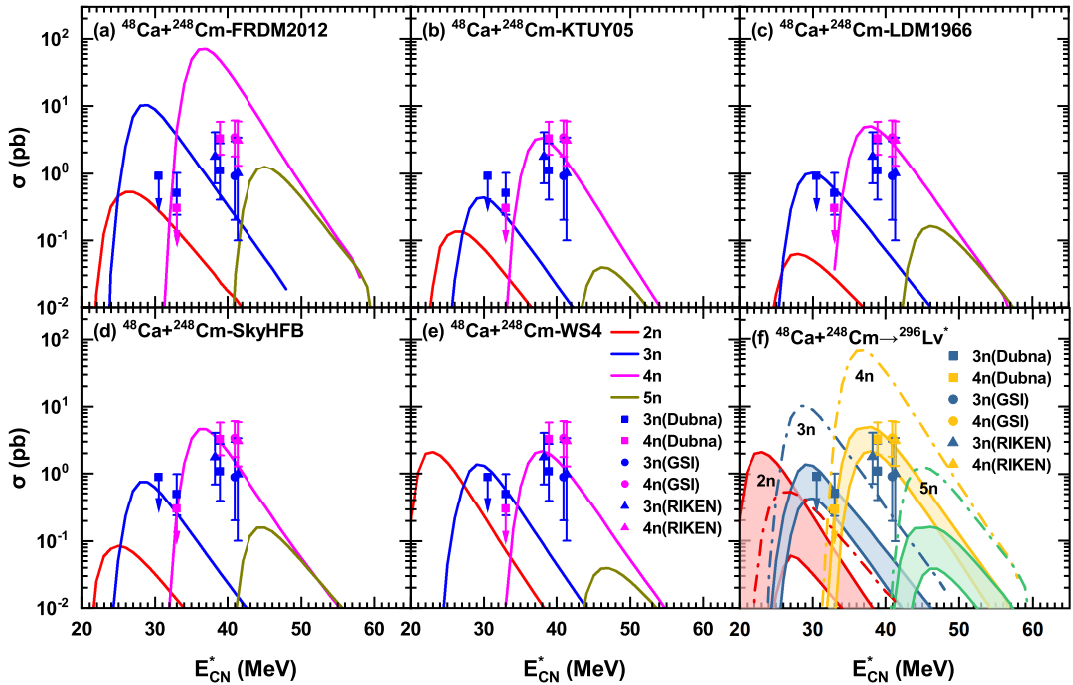}
\caption{The fusion-evaporation excitation functions in the reaction of $^{48}\mathrm{Ca} + ^{248}\mathrm{Cm}$ with different mass models. The experimental data of Dubna were taken from Ref. \cite{Yu2000, Yu2001, Yu2002, Yu2004, Yu2004(2)}, GSI data from Ref. \cite{Ho2012} and RIKEN data from Ref. \cite{Ka2017}. The arrow indicates the maximum cross-section estimated experimentally. The dotted lines in panel (f) represent the excitation functions provided by the FRDM2012 model, while the shaded areas are the excitation function ranges given by the other four mass tables.}
\end{figure*}

Fig. 2 shows the capture cross-section (a), fusion probability (b), and survival probability (c) of the reaction system $^{48}\mathrm{Ca} + ^{248}\mathrm{Cm}$ with different mass tables as a function of the excitation energy of the composite nucleus ($E^{\ast}_{CN} = E_{c.m.} + Q$). It is obvious that the survival probability rapidly decreases with the excitation energy of compound nucleus and the evaporating neutron number. The competition of fusion and survival probabilities dominates the optimal evaporation channel and excitation energy for producing the SHN. Shown in Fig. 3 is the excitation functions of this reaction system under different mass models.
The dot-dashed lines in Fig. 3 (f) represent the excitation functions provided by the FRDM2012 model, while the shaded areas are the excitation function ranges given by the left four mass tables. It can be seen that among the five mass models, for the 3n, 4n and 5n channels, the maximal cross-section values of FRDM2012 is the highest, about one order of magnitude higher than that of the other four mass tables. It is caused from the fact that the evaporation residual cross-section is determined by the direct product of the capture cross-section, the fusion probability and the survival probability. Firstly, the influence of different mass tables on the capture cross-section is not obvious, which can be seen in Fig. 2 (a). Secondly, it is obvious that when the excitation energy exceeds 30 MeV, the fusion probability given by the mass table FRDM2012 is the highest, followed by WS4, then LDM1966, and finally KTUY05 and SkyHFB. This explains why the excitation function given by the mass table FRDM2012 is generally higher than that of other mass models. From Fig. 2 (c), we can see that the survival probability of each evaporation channel of this system firstly increases and then rapidly decreases with the increase of excitation energy. The peak positions are respectively at 18$\sim$19 MeV (2n), 26$\sim$27 MeV (3n), 35$\sim$36 MeV (4n), and 43$\sim$45 MeV (5n), and the peak values vary by approximately one order of magnitude with different mass tables. It is worth noting that the survival probabilities given by the mass tables WS4, FRDM2012 and SkyHFB are often higher than those of the other two mass scales, which may be related to the neutron separation energy. Overall, from Fig. 3 (f), we can find that almost all the experimental data are in the shaded areas, indicating that to some extent, the DNS model considering cluster transfer is effective in studying the synthesis cross-section of superheavy nuclei. Therefore, based on these above five mass models, we extended the updated DNS model to other FE reactions induced by $^{48}$Ca and conducted a systematic comparison with the experimental data. Figure 4 and figure 5 describe the range of evaporation residue cross sections of 10 reaction systems for synthesizing SHE from darmstadtium (Z = 110) to oganesson (Z = 118) by bombarding actinide nuclei with the double magic nucleus $^{48}$Ca. The cross-sections were evaluated using five distinct mass tables. It is obvious that the magnitude of cross sections and structure of excitation functions sensitively depend on the nuclear mass models. Roughly, the 3n and 4n evaporation channels are available for synthesizing the SHNs and are consistent with the available experimental data from Dubna, RIKEN, Berkeley and GSI. The uncertainties of nuclear mass and fission barrier dominate the prediction of new superheavy nuclei. Overall, the experimental excitation functions in the FE reactions are nicely reproduced by the DNS model.

\begin{figure*}
\centering
\includegraphics[width=16 cm]{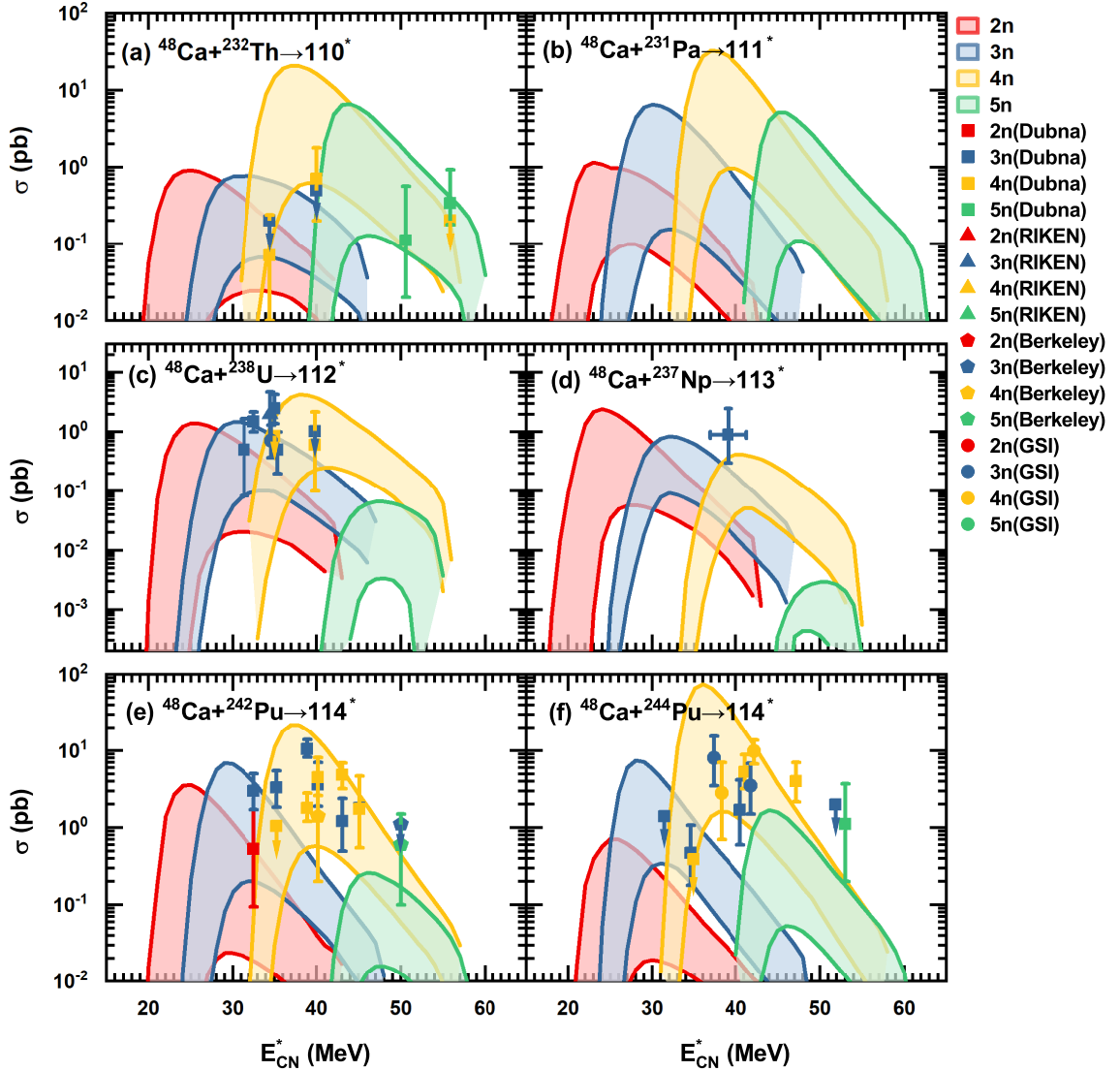}
\caption{ The excitation functions of the reaction systems of
$^{48}\mathrm{Ca} + ^{232}\mathrm{Th}$ \cite{Yu2023,Yu2024},
$^{48}\mathrm{Ca} + ^{231}\mathrm{Pa}$,
$^{48}\mathrm{Ca} + ^{238}\mathrm{U}$ \cite{Yu1999,Yu2004,Ho2007,Ka2017(2),Yu2022}, $^{48}\mathrm{Ca} + ^{237}\mathrm{Np}$ \cite{Yu2007},
$^{48}\mathrm{Ca} + ^{242}\mathrm{Pu}$ \cite{Yu2004,St2009,El2010,Yu2022},
$^{48}\mathrm{Ca} + ^{244}\mathrm{Pu}$ \cite{Yu2004,Yu1999(2),Yu2000(2),Yu2000(3),Yu2004(3),Yu2004(2),Du2010,Ga2011}  with different mass models. Symbols with arrows show upper cross-section limits. The shaded areas are the excitation function ranges given by the five mass tables. However, for $^{48}\mathrm{Ca} + ^{237}\mathrm{Np}$, since mass tables FRDM2012 and LDM1966 do not provide the excitation function information of 5n channels owing to the very low cross section.}
\end{figure*}

\begin{figure*}
\centering
\includegraphics[width=16 cm]{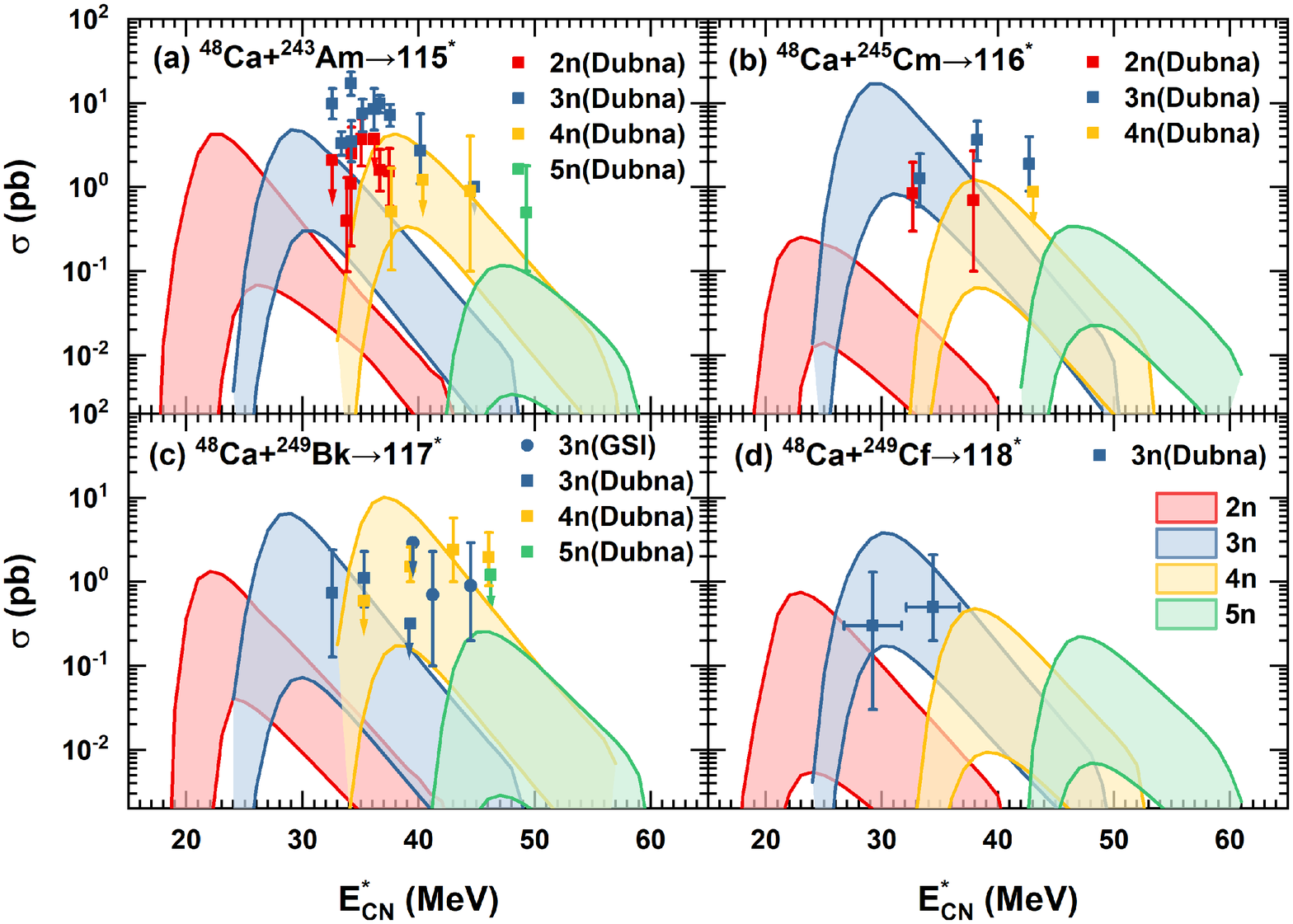}
\caption{The excitation functions of the reaction systems of $^{48}\mathrm{Ca} + ^{243}\mathrm{Am}$ \cite{Yu2004(4),Yu2005,Yu2012,Yu2013,Fo2016,Yu2022(2),Yu2022(3)}, $^{48}\mathrm{Ca} + ^{245}\mathrm{Cm}$ \cite{Yu2004(3),Yu2006}, $^{48}\mathrm{Ca} + ^{249}\mathrm{Bk}$ \cite{Yu2010,Yu2011,Yu2011(2),Yu2012(2),Yu2013(2),Kh2014,Yu2014,Kh2019}, $^{48}\mathrm{Ca} + ^{249}\mathrm{Cf}$ \cite{Yu2006,Yu2012(2)} with different mass models. The arrow indicates the maximum cross section estimated experimentally. The shaded areas are the excitation function ranges given by these five mass tables. }
\end{figure*}

In fact, most of the elements in the seventh period of the periodic table can be synthesized through fusion evaporation reactions between $^{48}$Ca and actinide atomic nuclei. However, the synthesis of elements beyond Z = 118 using $^{48}$Ca-induced FE reactions becomes increasingly challenging due to the difficulty of obtaining the desired target nuclei. Therefore, scientists began trying to synthesize SHN using FE reactions induced by other projectile nuclei. Recently, scientists in Berkeley synthesized the superheavy nucleus $^{290}$Lv for the first time using the $^{50}$Ti beam \cite{Ga2024}. Recently, Dubna also reported the experimental results on $^{50}\mathrm{Ti} + ^{242}\mathrm{Pu}$. The production cross-sections of the 3n and 4n channels are $0.32^{+0.34}_{-0.18}$pb and $0.22^{+0.27}_{-0.15}$pb with the excitation energy ($E^{\ast}_{CN}$) of 41 MeV, respectively. The cross section of $\sigma_{4n} = 32^{+46}_{-24}$fb at $E^{\ast}_{CN} = 42 MeV$ was measured for producing livermorium in the reaction of $^{54}\mathrm{Cr} + ^{238}\mathrm{U}$ \cite{Yu2025}. The experimental data pave a step stone for synthesizing new superheavy elements beyond oganesson. Furthermore, we used the DNS model to calculate the reaction system of $^{50}\mathrm{Ti} + ^{244}\mathrm{Pu}$. And the results based on the five mass tables are respectively presented in Fig. 6 (a), Fig. 6 (b), Fig.6 (c), Fig. 6 (d) and Fig. 6 (e). On the whole, 4n channel gives the largest cross section value, and the calculation results given by LDM1966 are most consistent with the experimental data \cite{Ga2024}. Fig. 7 predicts the range of the evaporation residual cross-sections of six reaction systems for the synthesis of SHE Z = 117$\sim$120 induced by $^{50}$Ti and $^{51}$V under five mass tables. As can be seen from the shaded area in Fig. 7, the maximum cross-sections of each evaporation channel given by these five mass tables fluctuate by approximately 1 to 2 orders of magnitude.

\begin{figure*}
\centering
\includegraphics[width=16 cm]{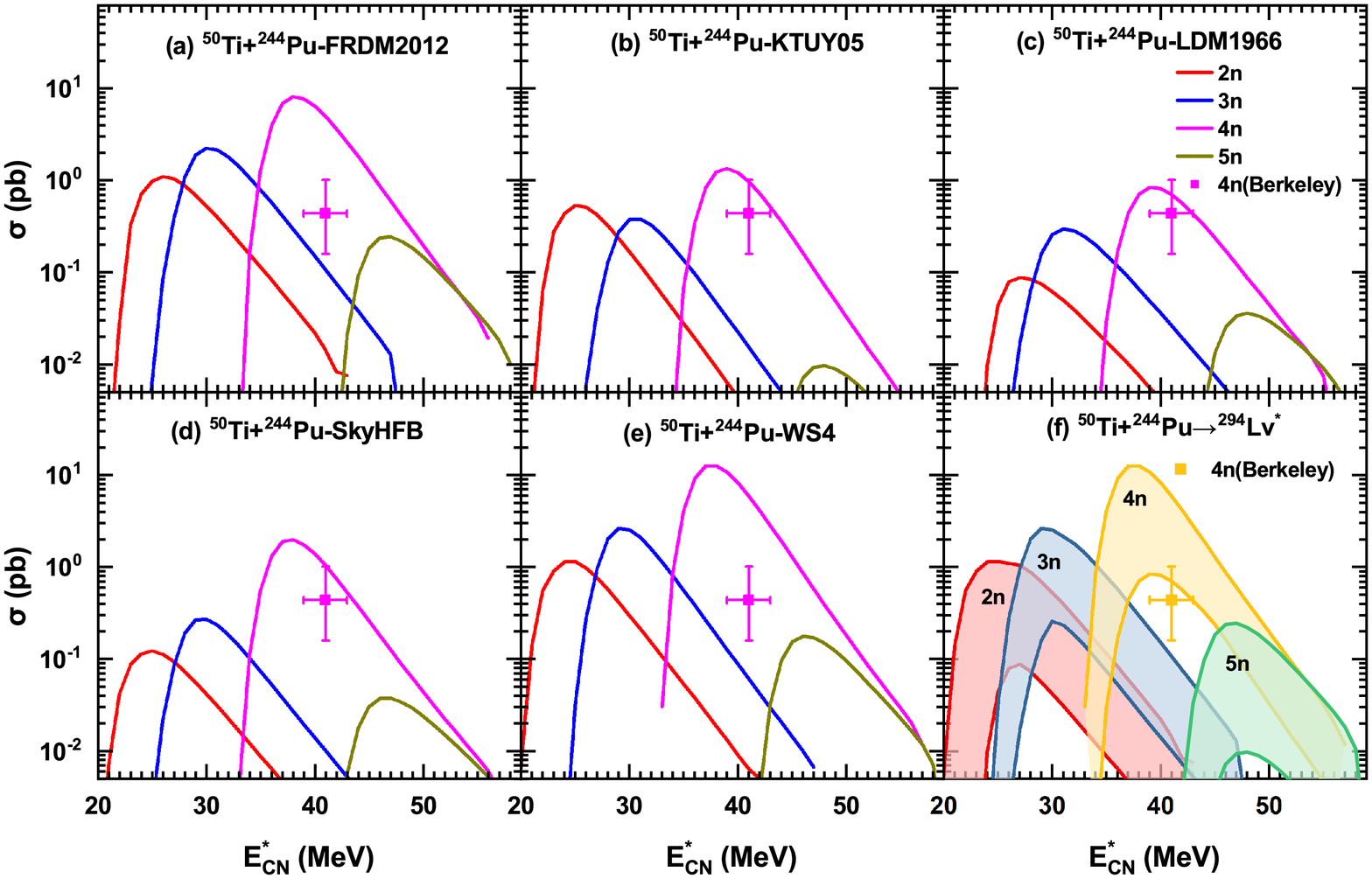}
\caption{Excitation functions of the reaction system of $^{50}\mathrm{Ti} + ^{244}\mathrm{Pu}$ \cite{Ga2024} with different mass models. The shaded areas are the excitation function ranges given by the five mass tables.}
\end{figure*}

\begin{figure*}
\centering
\includegraphics[width=16 cm]{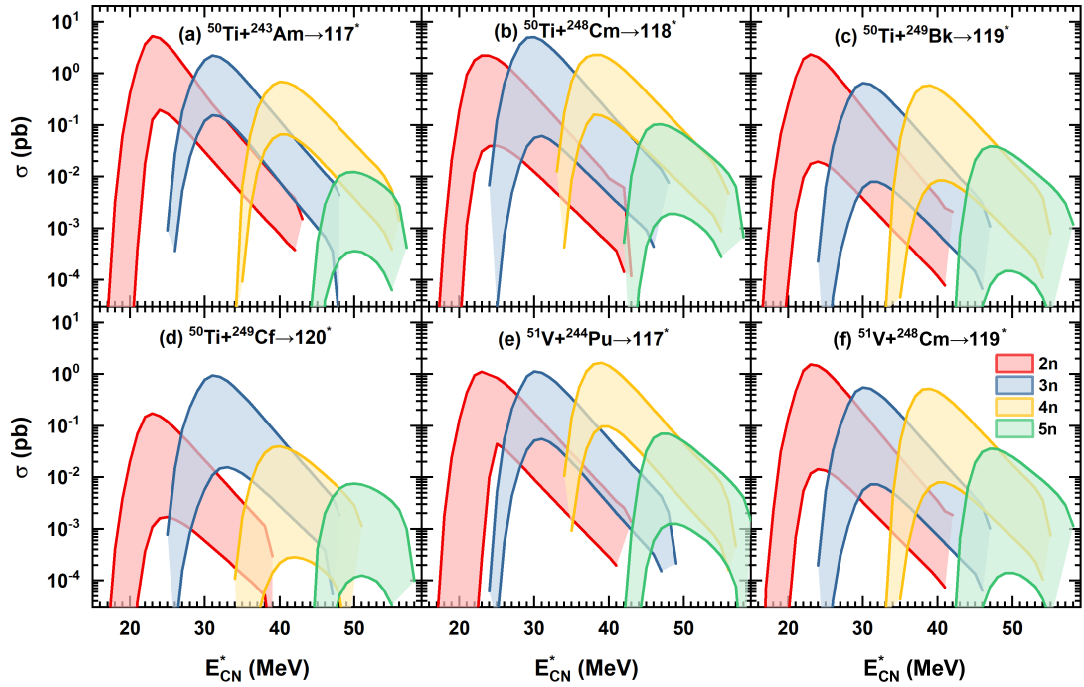}
\caption{The FE excitation functions for producing SHEs in the $^{50}$Ti and $^{51}$V induced reactions via the systems of $^{50}\mathrm{Ti} + ^{243}\mathrm{Am}/^{248}\mathrm{Cm}/^{249}\mathrm{Bk}/^{249}\mathrm{Cf}$ and $^{51}\mathrm{V} + ^{244}\mathrm{Pu}/^{248}\mathrm{Cm}$. }
\end{figure*}

\begin{figure*}
\centering
\includegraphics[width=16 cm]{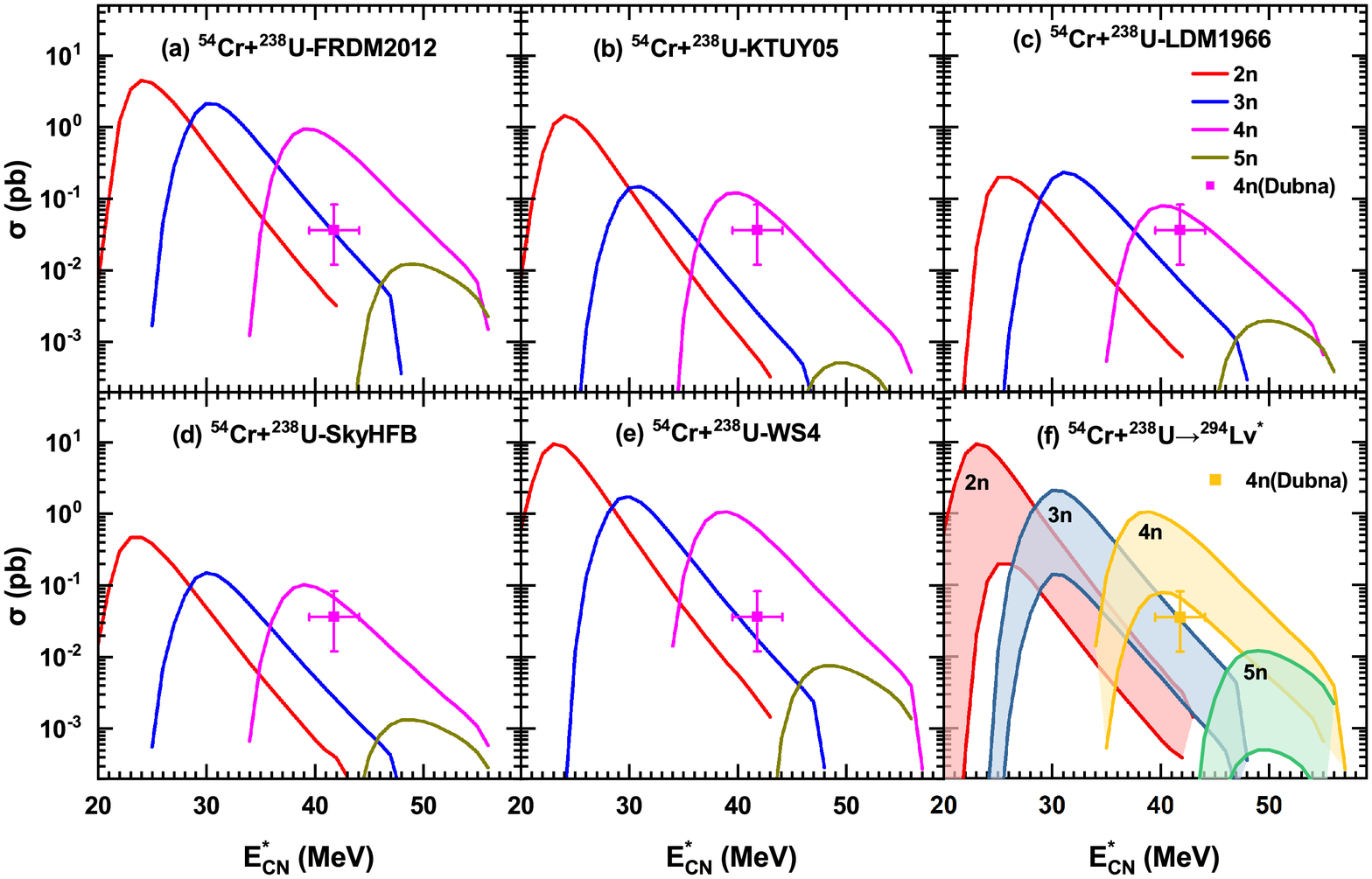}
\caption{The FE excitation functions for producing livermorium in the reaction of $^{54}$Cr + $^{238}$U and compared with the available data from Dubna \cite{Yu2025}. }
\end{figure*}

The $^{54}$Cr induced reactions on the actinide nuclei are a promising way to synthesize new SHEs. The feasibility of the $^{54}$Cr induced reactions and reaction mechanism are very valuable for the experimental measurements. Shown in Fig. 8 is the excitation functions in the reaction of $^{54}\mathrm{Cr} + ^{238}\mathrm{U}$ and compared with the recent data from Dubna \cite{Yu2025}. With the exception of the LDM1966 mass model, the left four models consistently demonstrate that, among the maximum cross-sections for producing livermorium, the 2n evaporation channel is available. The cross sections with the LDM1966 and SkyHFB mass models are consistent with the Dubna data. It should be noticed that the mass table and fission barrier are the key input quantities for predicting the production cross section of SHN.

\begin{figure*}
\centering
\includegraphics[width=16 cm]{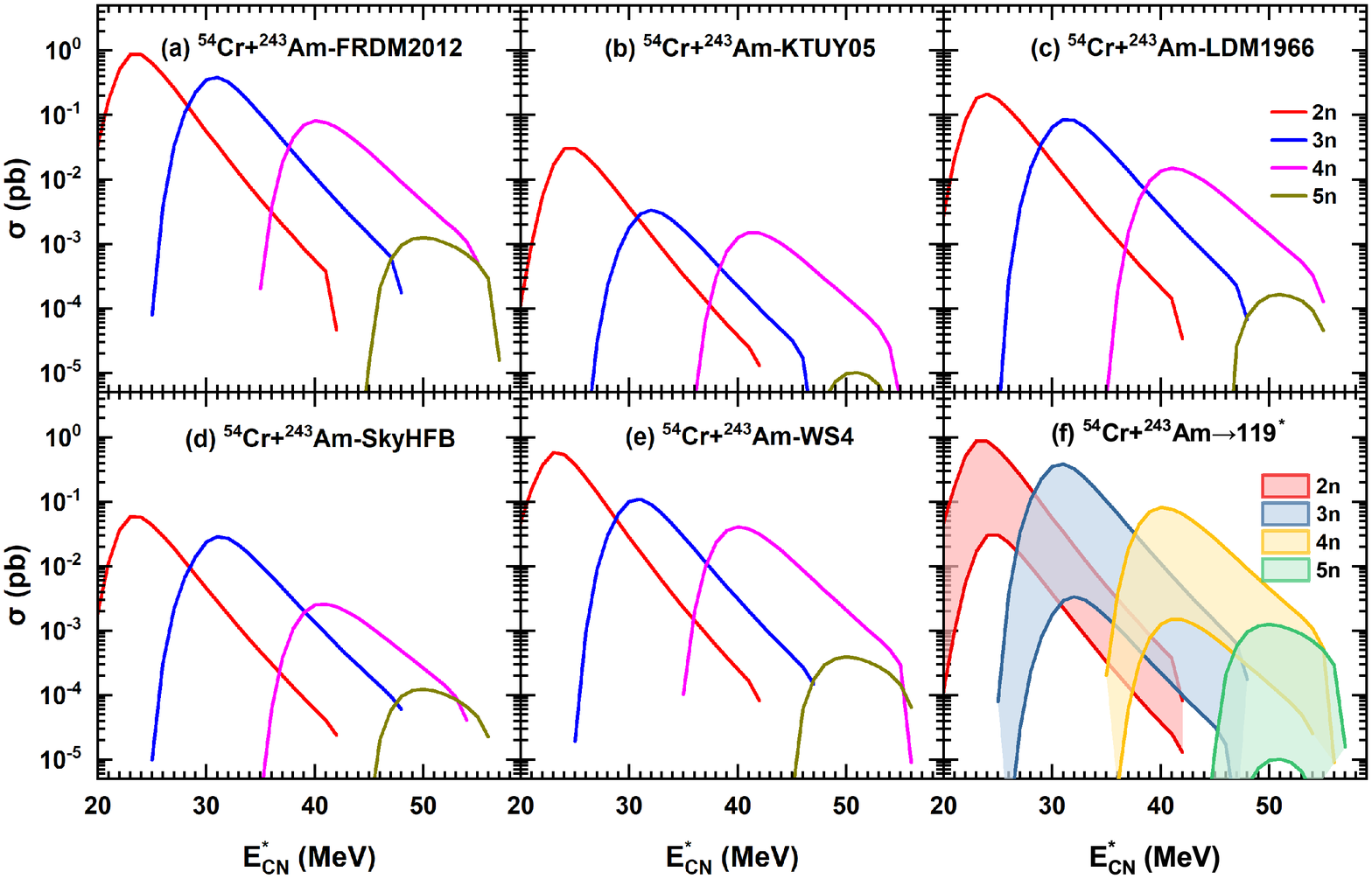}
\caption{The same as Fig.6, but for
$^{54}\mathrm{Cr} + ^{243}\mathrm{Am} \rightarrow 119^{*}$.}
\end{figure*}

\begin{figure*}
\centering
\includegraphics[width=16 cm]{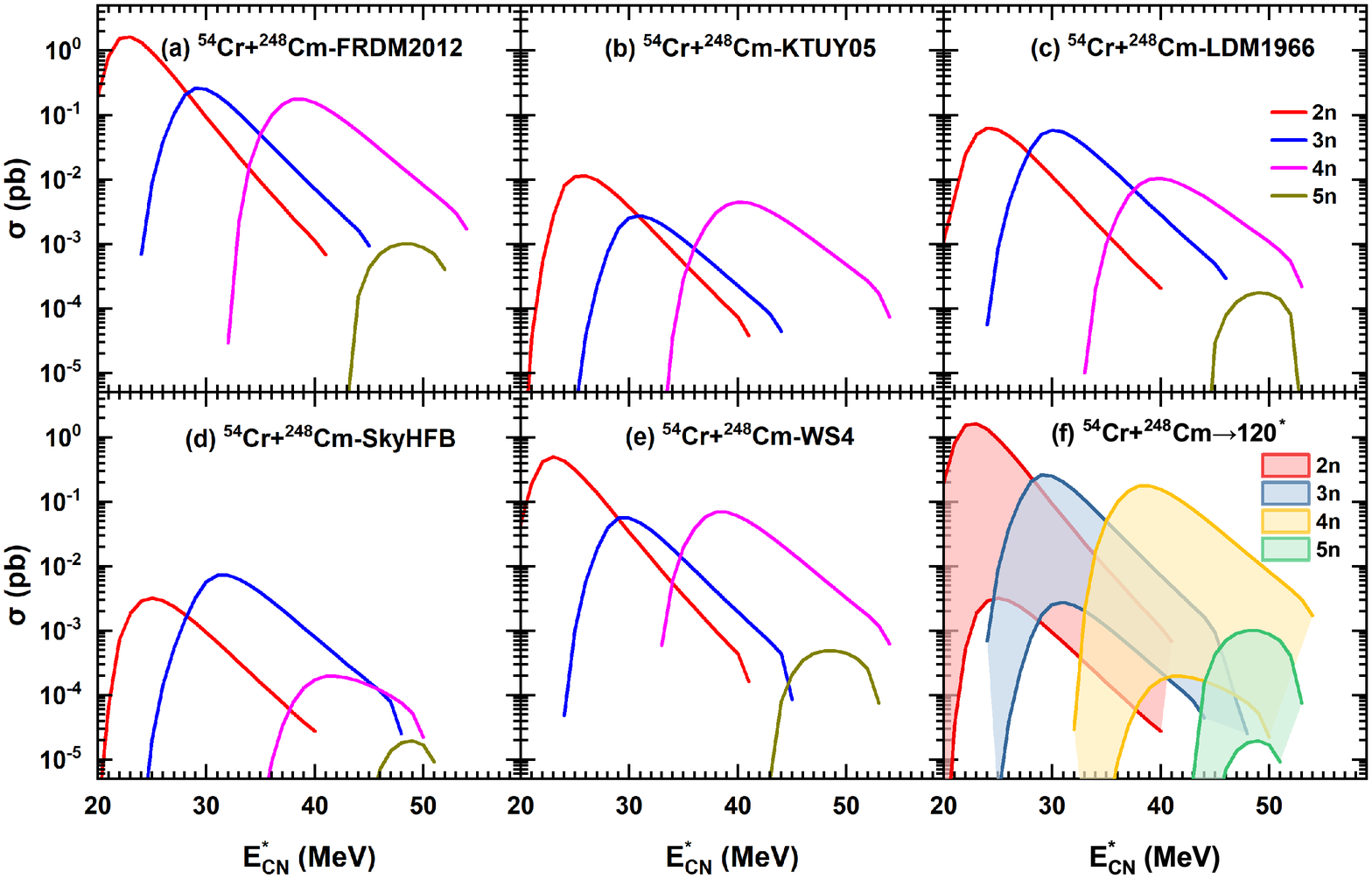}
\caption{The same as Fig.6, but for
$^{54}\mathrm{Cr} + ^{248}\mathrm{Cm} \rightarrow 120^{*}$. However, it is worth noting that since the KTUY05 mass table does not provide the excitation function information of 5n channels (in the program calculation, we ignored the output due to the small value), when calculating the range of the 5n excitation functions, we only considered the remaining four mass models.}
\end{figure*}

Based on the above analysis, we also calculated the excitation functions of $^{54}\mathrm{Cr} + ^{243}\mathrm{U} \rightarrow 119^{*}$ and $^{54}\mathrm{Cr} + ^{248}\mathrm{Cm} \rightarrow 120^{*}$, which are presented in Figures 9 and 10 for creating new elements Z=119 and 120. Among the five mass tables, FRDM2012 provides the largest cross section, followed by WS4, and KTUY05, which is related to the competition of fusion and survival probabilities. It is caused from the fact that, the mass table directly affects the calculation of the PES as shown in Eq. (20) during the fusion stage. The deexcitation process of the thermal SHN is determined by the neutron separation energy and fission barrier, which are estimated from the mass tables. It is obvious that the overall evaluation of the 2n evaporation channel is available for producing new SHE at the excitation energy around 25 MeV. The 2-4n channels are favorable for the new element production with the cross section above 1 fb. In Table II, we have summarized the maximum reaction cross-sections and corresponding excitation energies of 2n, 3n and 4n evaporation channels for synthesizing SHE Z = 119 and 120 under different mass tables. It should be mentioned that the 3n and 4n evaporation channels are the optimal way for the SHN production from darmstadtium (Z=110) to oganesson (Z=118) in the $^{48}$Ca induced reactions on the actinide nuclei. The $^{50}$Ti, $^{51}$V and $^{54}$Cr induced reactions enable the possibilities for creating the new elements beyond oganesson. The 2n channel cross section is enhanced because of the reduction of neutron separation energy for the compound nuclei of Z=119 and 120.

\begin{table*}
\centering
\caption{\label{Tab2} The maximum reaction cross sections and corresponding excitation energies of 2n, 3n and 4n evaporation channels for each reaction capable of synthesizing superheavy elements Z = 119 and 120 under different mass tables.}
\small
\begin{ruledtabular}
\begin{tabular}{cccccccccccccc}
\hline
& \multicolumn{2}{c}{FRDM2012} & \multicolumn{2}{c}{KTUY05} & \multicolumn{2}{c}{LDM1966} & \multicolumn{2}{c}{SkyHFB} & \multicolumn{2}{c}{WS4} \\
\hline
\cmidrule(r){2-3} \cmidrule(r){4-5} \cmidrule(r){6-7} \cmidrule(r){8-9} \cmidrule(r){10-11}
& $E^*/\text{MeV}$ & $\sigma_\text{max}/\text{pb}$ & $E^*/\text{MeV}$ & $\sigma_\text{max}/\text{pb}$ & $E^*/\text{MeV}$ & $\sigma_\text{max}/\text{pb}$ & $E^*/\text{MeV}$ & $\sigma_\text{max}/\text{pb}$ & $E^*/\text{MeV}$ & $\sigma_\text{max}/\text{pb}$ \\
\hline
\multicolumn{11}{c}{\textbf{$^{50}\mathrm{Ti} + ^{249}\mathrm{Bk} \rightarrow 119^{*}$}} \\
2n & 23.00 & 2.33E+00 & 24.00 & 3.46E-02 & 25.00 & 1.16E-01 & 24.00 & 1.97E-02 & 22.00 & 1.24E+00 \\
3n & 30.00 & 6.49E-01 & 31.00 & 7.86E-03 & 31.00 & 5.39E-02 & 31.00 & 3.01E-02 & 30.00 & 4.24E-01 \\
4n & 39.00 & 5.82E-01 & 40.00 & 8.44E-03 & 40.00 & 7.48E-02 & 39.00 & 2.87E-02 & 39.00 & 2.23E-01 \\
\midrule
\multicolumn{11}{c}{\textbf{$^{51}\mathrm{V} + ^{248}\mathrm{Cm} \rightarrow 119^{*}$}} \\
2n & 23.00 & 1.54E+00 & 25.00 & 2.58E-02 & 25.00 & 7.57E-02 & 24.00 & 1.43E-02 & 23.00 & 8.99E-01 \\
3n & 30.00 & 5.50E-01 & 31.00 & 7.21E-03 & 31.00 & 4.48E-02 & 31.00 & 2.72E-02 & 30.00 & 3.73E-01 \\
4n & 39.00 & 5.21E-01 & 40.00 & 7.88E-03 & 40.00 & 6.68E-02 & 39.00 & 2.71E-02 & 39.00 & 2.06E-01 \\
\midrule
\multicolumn{11}{c}{\textbf{$^{54}\mathrm{Cr} + ^{243}\mathrm{Am} \rightarrow 119^{*}$}} \\
2n & 23.00 & 8.80E-01 & 25.00 & 3.04E-02 & 24.00 & 2.11E-01 & 23.00 & 5.93E-02 & 23.00 & 5.83E-01 \\
3n & 31.00 & 3.85E-01 & 32.00 & 3.37E-03 & 31.00 & 8.55E-02 & 31.00 & 2.89E-02 & 31.00 & 1.09E-01 \\
4n & 40.00 & 8.15E-02 & 41.00 & 1.52E-03 & 41.00 & 1.52E-02 & 41.00 & 2.60E-03 & 40.00 & 4.08E-02 \\
\midrule
\multicolumn{11}{c}{\textbf{$^{50}\mathrm{Ti} + ^{249}\mathrm{Cf} \rightarrow 120^{*}$}} \\
2n & 23.00 & 1.71E-01 & 25.00 & 1.91E-03 & 25.00 & 1.08E-02 & 25.00 & 1.70E-03 & 23.00 & 1.22E-01 \\
3n & 31.00 & 9.39E-01 & 33.00 & 1.77E-02 & 33.00 & 2.55E-02 & 33.00 & 1.57E-02 & 31.00 & 5.29E-01 \\
4n & 40.00 & 4.00E-02 & 42.00 & 2.83E-04 & 41.00 & 4.08E-03 & 41.00 & 7.49E-04 & 40.00 & 1.17E-02 \\
\midrule
\multicolumn{11}{c}{\textbf{$^{51}\mathrm{V} + ^{249}\mathrm{Bk} \rightarrow 120^{*}$}} \\
2n & 24.00 & 2.82E+00 & 26.00 & 4.28E-02 & 25.00 & 7.42E-02 & 26.00 & 1.99E-03 & 24.00 & 1.09E+00 \\
3n & 31.00 & 2.18E-01 & 33.00 & 3.83E-03 & 33.00 & 2.75E-02 & 33.00 & 3.95E-03 & 31.00 & 1.09E-01 \\
4n & 40.00 & 2.79E-01 & 42.00 & 5.46E-03 & 42.00 & 1.15E-02 & 42.00 & 5.86E-03 & 40.00 & 1.31E-01 \\
\midrule
\multicolumn{11}{c}{\textbf{$^{54}\mathrm{Cr} + ^{248}\mathrm{Cm} \rightarrow 120^{*}$}} \\
2n & 23.00 & 1.64E+00 & 26.00 & 1.14E-02 & 24.00 & 6.28E-02 & 25.00 & 3.19E-03 & 23.00 & 4.96E-01 \\
3n & 29.00 & 2.61E-01 & 31.00 & 2.74E-03 & 30.00 & 5.86E-02 & 32.00 & 7.26E-03 & 30.00 & 5.60E-02 \\
4n & 38.00 & 1.75E-01 & 40.00 & 4.49E-03 & 40.00 & 1.04E-02 & 42.00 & 1.97E-04 & 38.00 & 6.88E-02 \\
\hline
\end{tabular}
\end{ruledtabular}
\end{table*}

\section{IV. Conclusions}
Within the framework of DNS model, the formation dynamics of SHN and the excitation functions in the FE reactions induced by $^{48}$Ca, $^{50}$Ti, $^{51}$V and $^{54}$Cr on actinide nuclei are systematically investigated and the calculations are based on the five mass models. It is demonstrated that the DNS model, which accounts for cluster transfer, exhibits excellent agreement with existing experimental data. Among the mass models considered, FRDM2012 generally yields the maximum cross-section values, followed by the WS4 mass model. The maximum cross-sections provided by the other three mass models do not exhibit significant differences. In general, the fusion-evaporation excitation functions given by the different mass models have the fluctuation about 1 or 2 orders of magnitude. Within the DNS model, the mass model directly impacts the calculation of the PES during the fusion process to form the compound nucleus. A lower internal fusion barrier is available for the nucleon or cluster transfer, in which the closer B.G. point to the incident position is formed, enhances the fusion probability in the competition with the quasifission process. The survival of excited SHN relies on the mass tables owing to the particle separation energies and the fission barrier. The available experimental data from Dubna, GSI, Berkeley and RIKEN are nicely reproduced with the improved DNS model for the SHN production from Cn (Z=112) to Og (Z=118) in the $^{48}$Ca fusion reactions on $^{238}$U, $^{237}$Np, $^{242,244}$Pu, $^{243}$Am, $^{245,248}$Cm, $^{249}$Bk and $^{249}$Cf. Furthermore, the FE reactions leading to the synthesis of new superheavy elements Z = 119 and Z = 120 induced by $^{50}$Ti, $^{51}$V and $^{54}$Cr on the target nuclides of $^{243}$Am, $^{248}$Cm, $^{249}$Bk and $^{249}$Cf have been systematically investigated. It is demonstrated that the channels of 2n and 3n evaporation are the optimal way to synthesize the new elements. More experiments are expected for the SHN measurements in the $^{50}$Ti, $^{51}$V and $^{54}$Cr induced reactions.

\section{Acknowledgements}
This work was supported by the National Natural Science Foundation of China (Projects No. 12575132, No. 12175072, No. W2412040, and No. 12311540139) and by the National Key Research and Development Program of China (Grant No. 2024YFE0110400).

\end{document}